# GENERATIVE AI AND AGENCY IN EDUCATION: A CRITICAL SCOPING REVIEW AND THEMATIC ANALYSIS

A PREPRINT

Jasper Roe [1*], Mike Perkins [2]

[1]James Cook University Singapore, Singapore.
[2] British University Vietnam, Vietnam.

[*] Corresponding Author: jasper.roe@jcu.edu.au

November 2024

## Abstract

This scoping review examines the relationship between Generative AI (GenAI) and agency in education, analyzing the literature available through the lens of Critical Digital Pedagogy. Following PRISMA-ScR guidelines, we collected 11 studies from academic databases focusing on both learner and teacher agency in GenAI-enabled environments. We conducted a GenAI-supported hybrid thematic analysis that revealed three key themes: Control in Digital Spaces, Variable Engagement and Access, and Changing Notions of Agency.

The findings suggest that while GenAI may enhance learner agency through personalization and support, it also risks exacerbating educational inequalities and diminishing learner autonomy in certain contexts. This review highlights gaps in the current research on GenAI's impact on agency. These findings have implications for educational policy and practice, suggesting the need for frameworks that promote equitable access while preserving learner agency in GenAI-enhanced educational environments.

***Keywords:*** *Generative AI, Learner Agency, Teacher Agency, AI in Education, Thematic Analysis, Critical Digital Pedagogy*

## Introduction

Generative Artificial Intelligence (GenAI) is still in its infancy, and understanding the use of GenAI in educational contexts is an immature field (Chiu, 2023). At the same time, a growing number of studies have begun to map out the dimensions in which GenAI may have an effect on education, and research agendas in some areas, such as tertiary education, have been initiated (Lodge, Thompson, et al., 2023). At the time of writing, there is no clear consensus as to whether GenAI tools offer a net positive or negative for education.

The current societal buzz surrounding AI contributes to the lack of clarity on exactly how these new technologies should be used in education. Media reports often highlight sensationalized impacts of the negative effects of GenAI technologies and predict disastrous societal consequences (Roe & Perkins, 2023). On the other hand, some research in educational contexts has shown positive effects on learning and teaching. For example, using a structured framework to enable the judicious use of GenAI tools in assessments can lead to fewer cases of academic misconduct and promote critical digital literacy (Furze et al., 2024; Perkins, Furze, et al., 2024), and the use of GenAI tools among college students may have a positive impact on academic achievement (Sun & Zhou, 2024). Furthermore, learners themselves are not necessarily averse to GenAI, with empirical studies demonstrating that learners in higher educational contexts are open to receiving GenAI feedback on assessment items if teacher oversight is present (Roe et al., 2024).

Other potential benefits of GenAI tools in education include providing personalized learning experiences, tutoring, and feedback (Kasneci et al., 2023), as well as allowing for the gamification of learning material and creation of digital resources (Zhai et al., 2021). Teacher collaboration with GenAI tools may help to reduce administrative burdens and develop effective learning resources for classrooms (Fui-Hoon Nah et al., 2023), or help with tasks such as tailoring curricula to local contexts (Karataş et al., 2024), and lessening the 'laborious tasks' of producing textual content for various educational purposes (Yan, Sha, et al., 2024). Furthermore, the use of a GenAI tool, ChatGPT, has been utilised for grading papers and has shown some measure of efficiency (Yavuz et al., 2024), albeit with some worrying tendencies, for example, avoiding very high or low scores, raising further ethical questions (Flodén, 2024).

In contrast to these benefits there are significant risks. GenAI tools are not transparent and produce unexplainable output (Miao & Holmes, 2023). Consequently, GenAI tools may be implemented in inequitable ways and contain unknown biases (Hacker et al., 2024; Kwak & Pardos, 2024). The risks of GenAI tools in education also relate to academic integrity and ownership of work (Cotton et al., 2023; Perkins, 2023; Rudolph et al., 2023), as well as the undetectability of GenAI outputs in submitted student work (Perkins et al., 2023; Perkins, Roe, et al., 2024; Chaka, 2024; Weber-Wulff et al., 2023). More broadly, societal impacts of GenAI in education include the possible pollution of the Internet with GenAI content, the worsening of digital poverty (Miao & Holmes, 2023), the 'monoculturing' of scientific knowledge (Messeri & Crockett, 2024), and the devaluing of human relationships (Resnick, 2024). GenAI outputs are also increasingly used for nefarious purposes (Ferrara, 2024), including in educational contexts, for example, in deepfake scandals or scams (Roe, Perkins, & Furze, 2024). In sum, the resulting impact of GenAI on education will not be known with clarity for some time, yet ongoing scholarship in this field can contribute to a fuller understanding of all the potential implications for students and teachers. From this perspective, we undertake a scoping review to explore the current body of literature on one significant topic: the relationship between agency and GenAI.

## GenAI and Agency in Education

There are multiple definitions of agency in relation to education, and although student agency is growing in importance as a field of study, it is commonly referred to without being clearly defined (Stenalt & Lassesen, 2022), and as a 'buzzword' rather than a developed notion (Inouye et al., 2023). At its simplest, agency may be seen as the capacity to make independent choices (Campbell, 2012) or the capacity to take intentional actions (McGivney, 2024). Educational research has suggested that learner agency can be developed (Ahn, 2016) and that learner agency is desirable as it is implicated in self-regulated learning, including self-efficacy, motivation, and metacognition (Xiao, 2014). Furthermore, environments that support the development of student agency are rich learning spaces for students and teachers (Vaughn, 2020).

It is important to note that while agency relates to independent choices, it is more than just an individual's capacity to act (Chisholm et al., 2019). Agency is situated contextually, temporally, and intra-personally (Manyukhina & Wyse, 2019), thus agency is emergent in the processes of social relations (Biesta et al., 2015). Moreover, learner agency involves the interplay between multiple constructs, such as abilities, affordance, self-regulatory systems, and motivation (Mercer, 2011) and developing agency can be considered from a process perspective, moving from powerlessness to a sense of control and future hopefulness (Blair, 2009). From a constructivist viewpoint, learner agency is necessary in the construction of knowledge and development of competence, requiring learners to voluntarily enter a Vygotskian Zone of Proximal Development (ZPD), and the teacher or instructor to equally enter into a scaffolding role; thus, learner agency may be co-constructed between teacher and student (Hempel-Jorgensen, 2015).

In other words, drawing on the above, it can be argued that agency is not a static phenomenon but develops and adapts in response to a system, varies across domains and contexts (Mercer, 2011), and is interlinked with our relationships with the external environment (Rappa & Tang, 2017); thus, agency can be 'shared' (Yang et al., 2020). From a critical perspective, agency is also embedded in power relations, for example, those that exist between teachers and students, or in the case of teacher agency, the autonomy offered by the environment in which they work (Rodriguez, 2013). Exercising agency can result in significant impacts. For Moje and Lewis (2007), who argue that learning is situated within a discourse community, agency has the potential to transform the self, personal identity, and relationships with others. In fact, as learning creates identities, it therefore positions learners in a way that they can 'take up' agency to alter dominant discourses, making agency central to learning (Moje & Lewis, 2007).

The limits and implications of agency are becoming a focal point when it comes to GenAI, with a major question being who takes responsibility for the output of GenAI-assisted or supported work. To this end, Dwivedi et al. (2023) point out that a consequence of increased AI use will be an increased focus on who is responsible in the case of errors, and if a GenAI tool is essentially an 'able slave' which we delegate our work to, we may ourselves become unable to write and think properly, thus impacting our own agency. In addition to this, the distinction between human agency and AI agency has attracted significant research attention. Dattahrani and De (2023) argue that in an AI-enabled world, the notion of agency must distinguish between AI, humans, and traditional information systems. Indeed, there is a sense that GenAI tools, such as ChatGPT, can seemingly exhibit agency or at least provide a convincing simulation of agency (Boulus-Rødje et al., 2024).

One of the key areas of discussion in this newly developing area of human-AI agency is in attempting to understand the extent to which GenAI systems may affect the agency of individuals in education, either positively or negatively. Research has suggested several potential benefits of GenAI technology in education on learner and teacher agency. Hwang and Chen (2023) argue that AI tools can guide learners based on their performance while enhancing their ability to understand, analyze, and assess. Lashari and Umrani (2023) emphasized that teachers should receive training in AI-assisted models to support student-centered learning, particularly for competence development and language acquisition. Lim et al. (2023) suggested that AI frameworks can be beneficial when they incorporate principles that promote learning agency, accuracy, and transparency.

It has also been argued that in new learning technologies such as Virtual Reality (VR), increased agency is an affordance, as it may enable full-body interaction within a digital environment (McGivney, 2024). A change in the learning environment through the use of GenAI may similarly lead to subsequent effects on agency in learners' perception of their learning process (Li et al., 2024), and AI-driven functionality may help increase learner agency in lifelong learning (Poquet & De Laat, 2021), with AI technologies specifically targeted towards promoting self-regulated and proactive learning in a digital context (Ng et al., 2024). In terms of theorized negative impacts, the automation of learning and education using AI technologies raises the concern that students' ability to engage in agentic decision-making will be limited (Darvishi et al., 2024), and there is the potential for GenAI tools, such as virtual teaching assistants, to disturb, intrude, or negatively impact learner agency (Resnick, 2024). In some cases, this may mean offloading challenging tasks, for instance, in language learning, to a GenAI tool, thus circumventing the work required to achieve a learning outcome (Roe et al., 2024). Given these varied ideas on how emerging technologies may enhance or diminish learner agency, it is necessary to examine these existing body of knowledge through a theoretical lens that can specifically address questions of agency in digital learning environments.

### Critical Digital Pedagogy as a Guiding Framework

In assessing the literature that currently exists regarding agency in learning and GenAI, we have chosen to draw on principles relating to the developing field of Critical Digital Pedagogy (CDP). CDP has yet to be fully defined, but broadly stems from the work of Freire's (1978) Pedagogy of the Oppressed. For Freire, all education belonged to the pursuit of freedom, which is by its nature political and focused on offering students a notion of critical agency. Freire rejected forms of pedagogy that reduced agency to consumer activity (Giroux, 2010). Following this, if students live in a culture in which their education takes place through digital tools such as screens, their education should seek to empower them in that context; thus, the critical and digital classroom is a context in which intellectual and moral agency is important (Rorabaugh, 2020). Furthermore, technology can be used to enable learners to make use of their agency or to allow learners to critically analyze technology tools in educational contexts (Masood & Haque, 2021). Simultaneously, educational technology has been criticized for its systems of surveillance, control, exploitation, and extraction of learners and teachers (Watters, 2020). Thus, when we speak about interpreting studies from a CDP perspective, we seek to understand whether such effects exist and explicitly identify these as areas of focus. Ultimately, the CDP is a relevant framework for this scoping study as it fundamentally focuses on agency - the agency to know, understand, act on, or resist reality (Morris & Jessifer, 2020). This lens helps us critically examine whether GenAI tools genuinely empower learners or potentially reinforce existing power structures in education.

## Materials and Methods

Our major objective for this scoping review is to systematically identify and analyze the existing literature regarding GenAI's impact on learner agency through the lens of CDP. This is a novel contribution to literature, as currently, little work has been done to map out the body of knowledge regarding this subject. Furthermore, we aim to identify gaps in the current research base and explore the potential for a future research agenda regarding GenAI and agency in an educational context. Our guiding questions for this research are as follows:

- How is the relationship between GenAI and learner agency conceptualized in academic literature?
- Through the lens of Critical Digital Pedagogy, what are the implications of GenAI for learner agency, considering issues of power, equity, and digital literacy?

We adopted a scoping review procedure under the principle that such reviews inform research agendas, map the concepts of a research area, and investigate the types of evidence available (Tricco et al., 2016). In our methodology, we followed the PRISMA Extension for Scoping Reviews (PRISMA-ScR) checklist developed by Tricco et al. (2018). We conducted a comprehensive search using four primary databases: Google Scholar, Scopus, Web of Science, and Education Resources Information Center (ERIC). These databases were chosen for their complementary strengths and coverage of the field. Table 1 provides a rationale for each database selected.

*Table 1: Database Selected*

| Database Selected for Scoping Review | Rationale |
|---|---|
| Scopus | SCOPUS contains extensive coverage of peer-reviewed literature in areas including computer science, education, and education technology. |
| Google Scholar | Scholar provides broad coverage across multiple disciplines and includes grey literature and preprints, which is particularly valuable given the rapidly evolving nature of Generative AI in education. |
| Web of Science (WoS) | Web of Science contains high-quality peer reviewed literature and allows us to discover studies that may have been missed in SCOPUS. |
| Education Resources Information Centre (ERIC) | Although many ERIC resources are listed in the above databases, ERIC specialises in education research and resources, ensuring thorough coverage of educational studies that may not have been listed in the above. |

The combination of these databases ensures that a wide net is cast over both interdisciplinary and education-specific research, capturing both established academic literature and emerging scholarship in the field of GenAI and agency in education.

The following search string was used, adapted for each database's specific syntax:

> ("Generative AI" OR "GenAI" OR "Large Language Model" OR "LLM" OR "ChatGPT") AND (education OR learning OR teaching) AND ("learner agency" OR "student agency" OR autonomy OR empowerment)

As shown in Table 2, strict inclusion criteria were set, and studies that did not meet these criteria were excluded. This included the exclusion of studies that focused solely on AI in education without addressing GenAI specifically, and those that did not substantially discuss learner or teacher agency, even if the term 'agency' was mentioned indirectly or as a single element in a larger study. We excluded all reports, such as news and media articles without a sufficient theoretical or academic focus, conference abstracts without full papers, and technical papers on GenAI that did not explicitly relate to agency and education.

*Table 2: Inclusion criteria*

| Inclusion Criteria | Detail |
|---|---|
| **Time of Publication** | Published at any time |
| **Language** | English |
| **Focus** | GenAI and Agency in Education |
| **Educational Level** | All levels of education |
| **Geographical Scope** | International (no geographical restrictions) |
| **Type of Publication** | Peer reviewed articles, conference papers, preprints, and relevant grey literature (e.g. reports from reputable organizations). |

**Screening Process**

Two reviewers independently screened the titles and abstracts of the identified studies against the inclusion and exclusion criteria. Full-text screening was conducted for potentially eligible studies. Given the limited number of studies that met our inclusion criteria, we did not employ formal inter-rater reliability (IRR) measures, such as Cohen's Kappa. Instead, we implemented a rigorous consensus-building approach in which both reviewers independently reviewed all papers and met to discuss each decision in detail. Disagreements were resolved through in-depth discussions and consultation of the inclusion/exclusion criteria. This approach, while less quantifiable than formal IRR measures, allowed for a detailed consideration of each study's relevance and content in the context of our research questions. A spreadsheet was used to extract relevant information from the included studies, including study characteristics, conceptualizations of learner agency, methodological approaches, and key findings, as well as noting any specific GenAI tools that were part of the research. Our initial database searches on Google Scholar, Scopus, Web of Science, and ERIC yielded 16 potentially relevant records. After reviewing these studies against our inclusion criteria, we excluded four studies for not being fully relevant to our research questions and one duplicate record. Ultimately, we identified only 11 studies in our qualitative synthesis, demonstrating the emerging and immature nature of the field in relation to learner agency and GenAI.

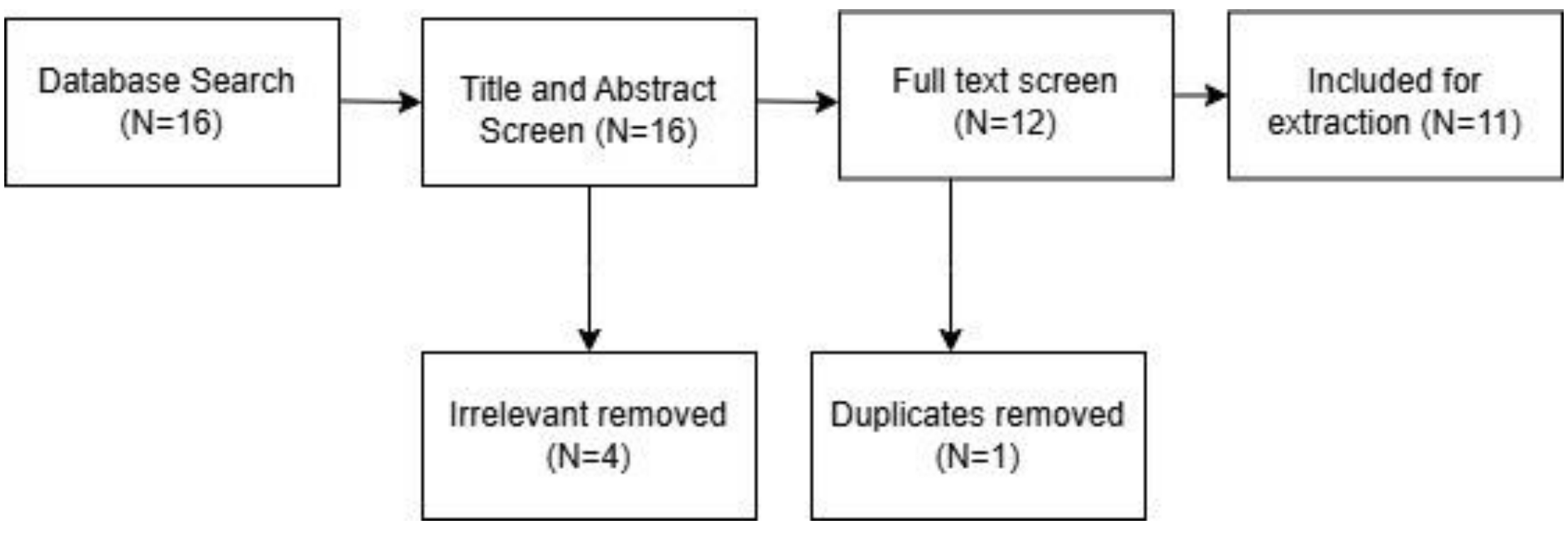

## Data Charting

Following the study selection process, we analyzed the characteristics of the 11 included documents. This section provides an overview of these studies, focusing on the key aspects relevant to our research questions on GenAI and learner agency.

All relevant studies were published in 2024, indicating the recent nature of research in this field. In total, nine studies focused on GenAI and learner agency in educational contexts, one focused on teacher agency, and one focused on broader notions of epistemic agency. Table 3 provides an overview of the key characteristics of these works. Most were published in peer-reviewed academic journals, with one conference paper and one preprint. The studies cover a range of educational contexts, including K-12, undergraduate education, doctoral education, language learning, and teachers (without a range or teaching level). In terms of methodologies, most studies were theoretical. The predominance of theoretical papers suggests that the field is still in its early stages, with researchers working to establish conceptual frameworks and theoretical understanding. However, the presence of empirical studies (participatory research, surveys, and experiments) indicates that some researchers are already gathering data on the practical implications of GenAI for agency in educational contexts.

*Table 3: Study Key Characteristics*

| Study | Author Details | Year | Type of Publication | Educational Context | Methodology |
|---|---|---|---|---|---|
| 1 | Kumar et al. | 2024 | Journal article | N/A | Theoretical paper |
| 2 | Godwin-Jones | 2024 | Journal article | Language Learning | Theoretical paper |
| 3 | Yang et al. | 2024 | Journal article | Undergraduate | Empirical study |
| 4 | Resnick | 2024 | Open access book | N/A | Theoretical paper |
| 5 | Shum | 2024 | Conference Proceedings | N/A | Theoretical paper |
| 6 | Yan et al. | 2024 | Conference Proceedings | Learning Analytics | Theoretical paper |
| 7 | Zhai | 2024 | Preprint | Teacher agency | Theoretical paper |
| 8 | Darvishi et al. | 2024 | Journal article | Undergraduate | Empirical study |
| 9 | Frøsig and Romero | 2024 | Preprint | Teacher agency | Theoretical paper |
| 10 | Han et al. | 2024 | Conference Proceedings | K-12 Students, teachers and parents | Empirical study |
| 11 | Cox | 2024 | Journal article | Epistemic agency | Theoretical paper |

### Hybrid Thematic Analysis

A hybrid thematic analysis methodology was employed to understand the data. Thematic analysis involves pattern recognition within the data, following an iterative, reflexive process (Fereday & Muir-Cochrane, 2006) and is often described in six steps, as detailed in Braun and Clarke's (2006) seminal work. However, recent work in thematic analysis has identified that using a hybrid approach, which combines inductive and deductive thematic analysis, can provide greater methodological rigor (Proudfoot, 2023). We followed this approach by first adopting a set of deductive codes based on the core principles of CDP as our guiding framework, while also working with the data to inductively make sense of the manuscripts, searching for common sets of meaning, and engaging in logical inference. This represents both a 'top-down' and 'bottom-up' approach to data analysis (Xu & Zammit, 2020). In line with current ideas regarding TA, we aimed to avoid generating mere 'topic summaries' and instead tried to creatively, organically, code the data to move beyond a superficial approach (Braun & Clarke, 2023).

To balance the deductive and inductive elements of the analysis, we used a systematic three-phase approach. This began with two researchers jointly developing a deductive framework using CDP principles to develop an initial high-level coding structure. This served as the analytical foundation for this study. We then conducted separate data familiarization and close reading of the manuscripts.

We opted to use GenAI tools to support code and theme triangulation. Previous research has shown that while GenAI tools may not be capable of reasoning in the manner of human coders, as statistical pattern-matching software, they can produce sophisticated analysis of data that may closely match the results of human researchers (Perkins & Roe, 2024a, 2024b), thus are suitable for supporting with triangulation of results. To achieve this, one researcher followed a traditional approach of taking memos during close reading of the source material, manually taking notes, and developing sub-codes for the high-level deductive codes. During this process, they also identified further inductive codes which described the 'qualitative richness of the phenomenon' (Boyatzis, 1998, p1) when encountering meaningful segments that did not align with our existing analytical framework. The second researcher used the frontier GenAI model Claude 3.5 Sonnet, produced by Anthropic, to analyze the source material and identify deductive and inductive codes and sub-codes, based on the agreed high-level coding structure.

In the third phase, we engaged in systematic comparison and discussion of the two sets of coding frameworks to identify any potential missing codes and sub-codes. Once an agreed set of codes and sub-codes was developed, a further step of theme development was carried out by both researchers (one supported by the GenAI tool). A final discussion and comparison of the themes helped resolve the differences between our findings and iteratively and reflexively refine our overall themes.

## Results

Our scoping review revealed both a complex and limited body of research regarding the role of GenAI in the agency of both learners and teachers. In all, despite a small number of eligible studies to review, we were able to develop several themes that raise critical questions about the future of learning in GenAI-enhanced environments from a CDP perspective. Our analysis resulted in the output of three themes that helped us interpret the data from the review: Control in Digital Spaces, Variable Engagement and Access, and Changing Notions of Agency. A summary of these themes and their descriptions is shown in Table 4.

*Table 4: Developed themes*

| Theme | Description | Example |
|---|---|---|
| **Control in Digital Spaces** | This theme refers to the negotiation of power and authority in GenAI-driven educational environments, in which traditional roles and mechanisms of control are being redefined. | 'There is an opportunity to consider HI (Hybrid Intelligence) to offer a potential answer in a system designed to give teachers (i) the power to act, (ii) the power to affect matters, (iii) the power to make decisions and choices, and (iv) the power to take a stance.' (Frøsig & Romero, 2024)' |
| **Variable Engagement and Access** | This theme covers patterns of varied engagement with GenAI in education, characterized by different levels of learner access, use and participation. | 'By analyzing student interviews conducted pre- and post-course, alongside their chat logs with GenAI and reflective journal entries detailing their learning approaches, the research uncovers a spectrum of student perspectives on GenAI's impact, ranging from beneficial optimism, to cautious skepticism and adaptable pragmatism.' (Yang et al., 2024)' |
| **Changing Notions of Agency and Human-AI Hybrids** | This theme relates to the changing ways in which learners incorporate critical capabilities and strategic approaches to using GenAI in education. | 'For example, some students may use GenAI tools to improve the grammar and presentation of their reflective writing, while others may ask these tools to generate the entire reflective assessment, undermining the educational purpose of such tasks' (Yan, Martinez-Maldonado, et al., 2024)' |

## Control in Digital Spaces

This theme is related to the undertaking of action in digital spaces and the way in which learners make agentic decisions on how to use GenAI. We noted that within the literature, there was a strong focus on how specific learning actions, such as submitting work, drafting, or reviewing, may be engaged by learners and why. As an example, Yan et al.'s (2024) framework sought to understand how students exert control when using GenAI tools and noted that learners exhibit different behaviors, including whether to reject the use of such tools or integrate them with varying methods and techniques. Those who used the tools most effectively sat on the 'reflective' end of the spectrum and were able to demonstrate critical skills such as introspection, reflection, and evaluation. Such an approach aligns with Bearman et al.'s (2024) notion of evaluative judgement as an essential tool for the future.

Parker et al. (2024) equally discusses the importance of selective use of GenAI outputs, meaning that learners exhibit both agency and CDL skills in using GenAI. This reflects the focus of CDP, in which learners are co-creators of knowledge rather than passive consumers. On the other hand, we noted little discussion of the potential constraints that GenAI tools may place on learners' agency. It can be asked whether the limitations of GenAI tools, for example their culture-bound knowledge (Roe, 2024) or potential for manipulation (Ferrara, 2024), create conditions that constrain the agency of the learner and limit their choices for engaging in academic tasks.

## Variable Engagement and Access

The potential for GenAI to exacerbate educational inequalities is a concern that emerged during our review and has been noted in the literature. This aligns with broader thinking on the relationship between AI and education, which postulates that unequal access may worsen digital poverty and drive a divide between those with and those without (Miao & Holmes, 2023). In our review, we found data to support the awareness that limited access to GenAI tools may constrain learner agency, but also that resistance to the use of GenAI is possible. Shum (2024), for example, warns that the disadvantages faced by those unable or unwilling to use AI tools highlight the risk of creating new forms of digital divide, while Frøsig and Romero (2024) also highlight that in the domain of teacher agency, it is important to recognize that GenAI may widen the digital divide among those students who can and cannot afford access to AI tools.

Further evidence underpinning this theme comes from Yan et al. (2024), who envision a world in which affluent learners may experience more dynamic, personalized learning content than less affluent learners, while on the other hand, Godwin-Jones (2022) argues the opposite, framing AI tools as a way to reduce social and linguistic inequalities, for example, those with a lower level of English proficiency. Therefore, the extent to which GenAI may impact agency by enhancing or exacerbating educational inequalities is a topic of clear interest in the literature, and led us to the development of this theme. From a CDP perspective, this theme strongly relates to questions of equality, power, and the potential social effects of GenAI on learner agency.

## Changing Notions of Agency and Human-AI Hybrids

The third theme we developed within our scoping review is based on the concept of agency in GenAI-enhanced learning environments undergoing a shift. This means that agency is seen as changing to encompass nonhuman GenAI technologies. The idea that agency is 'shared' (Yang et al., 2020) is not new, and agency is related to the external environment too (Rappa & Tang, 2017). Yet, agency in a period of GenAI may now be shared between humans and non-human technologies, as proposed by Godwin-Jones (2022). Similarly, as Dattahrani and De (2023) contend, agency must now be shared between AI actors and humans, and AI tools such as ChatGPT may be able to exhibit agency (or at least appear to) (Boulus-Rødje et al., 2024). From a CDP perspective, this provokes questions as to whether individuals should resist the forfeiture of agency to machinery.

With this in mind, more work is needed to describe exactly what level of agency we may ascribe to GenAI tools, and whether they have some semblance of agentic behavior or are merely an 'able slave' (Dwivedi et al., 2023). From a CDP perspective, reconceptualizing notions of agency and to whom it belongs is part of a broader examination of the power relationships inherent in learning environments, particularly those which are digitally enhanced. This aligns with Cox's (2024) analysis of evolving epistemic agency in AI-enhanced education, where he argues that traditional notions of learners as 'makers' of knowledge are being challenged by new conceptualizations of learners as 'managers' of information or even as 'information organisms' (inforgs) embedded within AI systems. If GenAI tools are being used not just as tools, but as more active participants in the learning process and knowledge construction, then we see a greater opportunity for the types of 'entanglement' between human and non-human actors as described by Kumar (2024). This suggests that new frameworks must be developed to understand exactly how agency operates in hybrid human-AI collaborative learning environments. These must account for both the algorithmic decision-making principles of GenAI tools, as well as the intentional actions of human educators and learners.

## Discussion

Our scoping review of the literature on GenAI and learner agency in education, viewed through the lens of Critical Digital Pedagogy (CDP) resulted in the development of three themes that carry implications for future research in this area.

The focus on learner control and the potential to empower or diminish learner agency has also recurred throughout the literature. For example, Resnick (2024) stated that AI tutors may disempower or intrude on the learning process. From a CDP perspective, this warrants additional attention – the product-focused nature of many AI-driven EdTech tools, and the drive for profit means that any benefits to learning are likely to be emphasized, while this possibility of disempowerment is likely to be downplayed on the commercial stage. For this reason, it is vital to conduct impartial, independent studies on whether such tools have an empowering, disempowering, or mixed effect on learner agency.

A further finding that emerged from our review is the way in which GenAI may contribute to reshaping knowledge ownership and construction in educational environments. In Han et al. 's (2024) study, for example, the authors observed that among 8–12-year-old students, when engaging with GenAI in a storytelling task, many simply copied and pasted outputs directly from the GenAI tool without making any amendments. In relation to agency and the intellectual engagement and autonomy that CDP advocates strive for, this is a concerning issue. However, why learners do this and whether this represents a diminishment of agency requires further research. Our review also highlighted that existing studies acknowledged and discussed the potential for GenAI to exacerbate existing societal and educational inequalities. Shum (2024) for example, warned that professionals unable or unwilling to use AI tools might find it impossible to meet the productivity of those who embrace AI tools, thus creating an inequitable environment for those with aptitude and access and those without. On the other hand, some studies noted GenAI's potential for increased personalization, which could align with CDP's emphasis on learner-centered education if implemented critically. The need for ethical frameworks and responsible AI use in education was also a recurring pattern in the data. Parker et al. (2024) stressed the need for developing frameworks for responsible AI use in learning environments, while Yan et al. (2024) highlighted the importance of transparency in AI use, aligning with CDP's emphasis on demystifying digital tools and processes.

Our analysis revealed at times a critique of techno-deterministic views of GenAI in education, resonating with CDP's skepticism towards technological solutionism. Kumar (2024) challenged the notion of GenAI inevitably transforming education, emphasizing that human lives are inevitably entangled in multiple non-human learning-related tools. This perspective aligns with CDP's view of technology as entangled with social and cultural contexts. These findings demonstrate a developing picture of GenAI for learner agency in education. Viewed through the critical lens of CDP, there is clearly an urgent need for critical engagement with these technologies, the development of refined digital literacies, and careful consideration of ethical implications. Simultaneously, the unclear balance between the potential benefits and risks of GenAI presents significant challenges for researchers, educators, and policymakers.

\

## Limitations

While this scoping review provides insights into the intersection of GenAI and learner and teacher agency in education, there are several limitations that affect the interpretation and replicability of our findings. The field of GenAI in education is rapidly evolving, with all the included studies published in 2024, suggesting that this topic is only just beginning to be explored. Consequently, long-term impacts and trends are not yet observable. Furthermore, the capabilities of GenAI tools are advancing at such a speed that any potential impact on agency will likely shift as new capabilities and forms of GenAI become available. The use of GenAI tools (Claude 3.5 Sonnet) in our analysis process, while aimed at enhancing the identification of sub-codes and support in the creation of themes, may have introduced unforeseen biases or limitations in the interpretation of data. The capabilities and limitations of the AI model used should be considered when interpreting our results.

A significant limitation is also the dominance of theoretical studies in the identified data. While these provide interesting insights, discussions, and considerations of how agency might shift in relation to GenAI, the lack of practical applications and empirical studies makes it difficult to measure the success of theoretical models against real-world applications. Finally, our review included a small number of studies, which constrains how broad and deep our analysis could be. That said, in identifying the key issues currently underpinning this topic, the review captures a snapshot of the current concerns regarding GenAI and agency in education. Given the recent emergence of GenAI technologies in education, our review lacks a longitudinal perspective of the evolution of the relationship between GenAI and learner agency over time.

While Critical Digital Pedagogy provides a valuable theoretical lens, operationalizing its principles in the context of GenAI and learner agency presents challenges. The interpretation and application of CDP concepts may vary among researchers. As with any review, there is also a risk of publication bias, as studies with positive or significant findings are more likely to be published and included in our review. Given the theoretical nature of the majority of the papers identified, this is unlikely to be a significant limitation.

## Conclusion

This scoping review reveals that, at present, studies on learner agency and GenAI focus on issues of control, access, engagement, and exploring a shifting notion of agency. Viewed through the lens of CDP, our findings suggest that further research is required on the power dynamics at play when GenAI is implemented with a view to enhancing learner agency, and we encourage critical research that questions the assumption that technology will by its nature enhance learning and improve the agency of learners and teachers. Such research should aim to address questions of cultural bias and knowledge of GenAI models, and how they may constrain, rather than enhance learner agency and freedom to learn, or focus on ways that GenAI may contribute to diminishing the agency of learners or exacerbate inequalities.

**Data availability statement**
N/A

**Funding statement**
No funding was received for this study.

**Conflict of interest disclosure**
The authors have no conflicts of interest to declare.

**Ethics approval statement**
N/A

**Patient consent statement**
N/A

**Permission to reproduce material from other sources**
N/A

**Clinical trial registration**
N/A